\documentstyle[12pt,twoside,fleqn,espcrc1,epsf]{article}
\newcommand{\beq}{\begin{equation}}
\newcommand{\eeq}{\end{equation}}
\newcommand{\beqa}{\begin{eqnarray}}
\newcommand{\eeqa}{\end{eqnarray}}

\newcommand{\AmS}{{\protect\the\textfont2
  A\kern-.1667em\lower.5ex\hbox{M}\kern-.125emS}}

\title{Nucleon form factors: From the space--like to the time--like region} 
\author{Ulf-G. Mei{\ss}ner\address{Forschungszentrum J\"ulich, 
Institut f\"ur Kernphysik (Theorie), 
D-52425 J\"ulich, Germany}}
\begin{document}

\maketitle

\begin{abstract}
\noindent I discuss how  dispersion relations can be used to  analyse
the nucleon electromagnetic form factors, with particular
emphasis on the constraints from unitarity and pQCD. Results for
nucleon radii, vector-meson couplings, the onset of pQCD and bounds  
on the strangeness form factors are presented. The em  form
factors in the time--like region reveal some interesting physics
which is not yet understood in full detail.
 The need for a better data basis at low, intermediate and large
momentum transfer and also in the time--like region is stressed. 
\end{abstract}

\section{Objectives}

\noindent There are many reasons to analyse the nucleon electromagnetic
(em) form factors (ffs) with a precise theoretical tool, 
in this case dispersion relations:

\begin{itemize}

\item[$\star$] {\it Nucleon radii}:  The structure of the nucleon in the
non--perturbative regime is characterized by certain scale parameters,
like the magnetic moments, em radii, em polarizabilities and so on. 
Since the form factors are essentially Fourier-transforms of the charge
and magnetization distribution in the nucleon, they give information on the
first moments, i.e. the em radii. A precise knowledge of these quantities
is not only interesting {\it per se}, but also is mandatory to further reduce
the theoretical uncertainty in the Lamb shift analysis which serves as an
excellent high precision test of QED \cite{lamb}.   

\item[$\star$] {\it Coupling constants}: It is well known that the
spectral functions of the em form factors contain information about 
the vector-meson--nucleon coupling constants, in particular the 
tensor-to-vector coupling ratio. Remember that the vector mesons were
indeed predicted from studies of electron--nucleon scattering data
before they were actually detected in pion--nucleon reactions  
\cite{frafu}\cite{na}\cite{sa}. 

\item[$\star$] {\it Strangeness in the nucleon}: The spectral functions
contain also information about the much discussed strange vector form 
factors, which are and will be measured at SAMPLE, TJNAF and MAMI. As I
will show, there is still very active debate how to extract such information
either via $\phi$-meson couplings and/or kaon loops and to what extend such
estimates are reliable. 

\item[$\star$] {\it Onset of pQCD:} Perturbative QCD (pQCD) tells us 
how the nucleons ffs behave at very large momentum transfer 
(in the space-- and the time--like region) based on dimensional counting
arguments supplemented with the leading logs due to QCD \cite{pQCD}. 
The dispersive analysis can shed some light on were the onset of 
QCD scaling could  be expected.

\item[$\star$] {\it Mystery of the time--like ffs}: The recent analysis
of the FENICE group of the ADONE data for the nucleon ffs in the time--like
region hints at some interesting structure just below the two nucleon
threshold (isovector resonance, dibaryon, $\ldots$?) 
\cite{fen}. To understand these
data consistently with the space--like ones is a major challenge.

\end{itemize}

In the following, I will show how one can address these questions in the
framework of dispersion relations and discuss the pertinent results. 

\section{The tool: Dispersion relations}

\noindent The structure of the nucleon (denoted by '$N$')
as probed with virtual photons
is parametrized in terms of four form factors,
\beq
<N(p')\, | \, {\cal J}_\mu \,  | \, N(p)> 
= e \,  \bar{u}(p') \, \biggl\{  \gamma_\mu F_1^{N} (t)
+ \frac{i \sigma_{\mu \nu} k^\nu}{2 m_N} F_2^{N} (t) \biggr\} 
\,  u(p) \,, \quad N=p,n \,,
\eeq
with $t = k_\mu k^\mu = (p'-p)^2$ the invariant momentum 
transfer squared, ${\cal J}_\mu$ 
the em current related to the photon field and $m_N$ the nucleon mass.
In electron scattering, $t < 0$ and it is thus convenient 
to define the positive
quantity $Q^2 = -t > 0$. $F_1$ and $F_2$ are called the Pauli and the Dirac
form factor, respectively, with the normalizations $F_1^p (0) =1$,
$F_1^n (0) =0$, $F_2^p (0) =\kappa_p$ and $F_2^n (0) =\kappa_n$. Here,
$\kappa$ denotes the anomalous magnetic moment.
Also used are the electric and magnetic
Sachs ffs,
\beq
G_E = F_1 - \tau F_2 \, , \quad G_M = F_1 + F_2 \, ,
 \quad \tau =Q^2/4m_N^2 \,.
\eeq
In the Breit--frame, $G_E$ and $G_M$ are nothing but the Fourier--transforms
of the charge and the magnetization distribution, respectively.
There exists already a large body of data
for the proton and also for the neutron. 
In the latter case, one has to perfrom
some model--dependent extractions to go from the deuteron
or $^3$He to the neutron. 
More accurate data are soon coming (ELSA, MAMI, TJNAF, $\dots$). 
There are also data in the time--like region from the reactions $e^+ e^-
\to p \bar{p} , n \bar{n}$ and from annihilation $p \bar{p} \to
e^+ e^-$, for $t \ge 4m_N^2$. It is
thus mandatory to have a method which allows to analyse all these data in a
mostly model--independent fashion. That's were dispersion theory comes into
play. Although 
not proven strictly (but shown to hold in all orders in perturbation
theory), one writes down an unsubtracted dispersion relation for $F(t)$ (which
is a generic symbol for any one of the four ff's),
\beq
F(t) = \frac{1}{\pi} \int_{t_0}^\infty \, dt' \, \frac{{\rm Im} \, F(t)}{t'-t}
\, \, , \eeq
with $t_0$ the two (three) pion threshold for the isovector (isoscalar) ffs.
Im~$F(t)$ is called the {\it spectral}
{\it  function}. It is advantageous to work in
the isospin basis, $F_i^{s,v} = (F_i^p \pm F_i^n)/2$, since the photon
has an isoscalar ($I=s$) and an isovector ($I=v$) component. These spectral 
functions are the natural meeting ground for theory and experiment, 
like e.g. the partial
wave amplitudes in $\pi N$ scattering. In general, the spectral functions 
can be thought  of as a superposition of vector meson poles and some 
continua, related to n-particle thresholds, like
e.g. $2\pi$, $3\pi$, $K \bar{K}$, $N\bar{N}$ and so on. 
For example, in the Vector Meson
Dominance (VMD) picture one simply retains a set of poles.  If the data were 
to be infinitely precise, the continuation from negative $t$ (data) 
to positive $t$ (spectral functions)
in the complex--$t$ plane would lead to a unique result for the
spectral functions. Since that is not the case, one has to make some extra
assumption guided by physics to overcome the ensuing instability as will be
discussed below. Let me first enumerate the various constraints one has 
for the spectral functions.

\section{Constraining the spectral functions}

\noindent It is important to realize that  there are some
powerful constraints which the spectral functions have to obey. Needless
to say that in many models of the nucleon em ffs, only some of these
are fulfilled. Consider first the spectral functions just above threshold.
Here, {\it unitarity} plays a central role. As pointed out by Frazer and 
Fulco  \cite{frafu}
long time ago, extended unitarity leads to a drastic enhancement 
of the isovector spectral functions on the left wing of the $\rho$ resonance.
Leaving out this
contribution from the two--pion cut leads to a gross underestimation of the
isovector charge and magnetic radii. This very fundamental constraint is very
often overlooked. In the framework of chiral pertubation theory, this 
enhancement is also present at the one--loop level as first shown in
ref.\cite{gss}. 
\begin{figure}[ht]

\hskip 4in
\epsfysize=2.2in
\epsffile{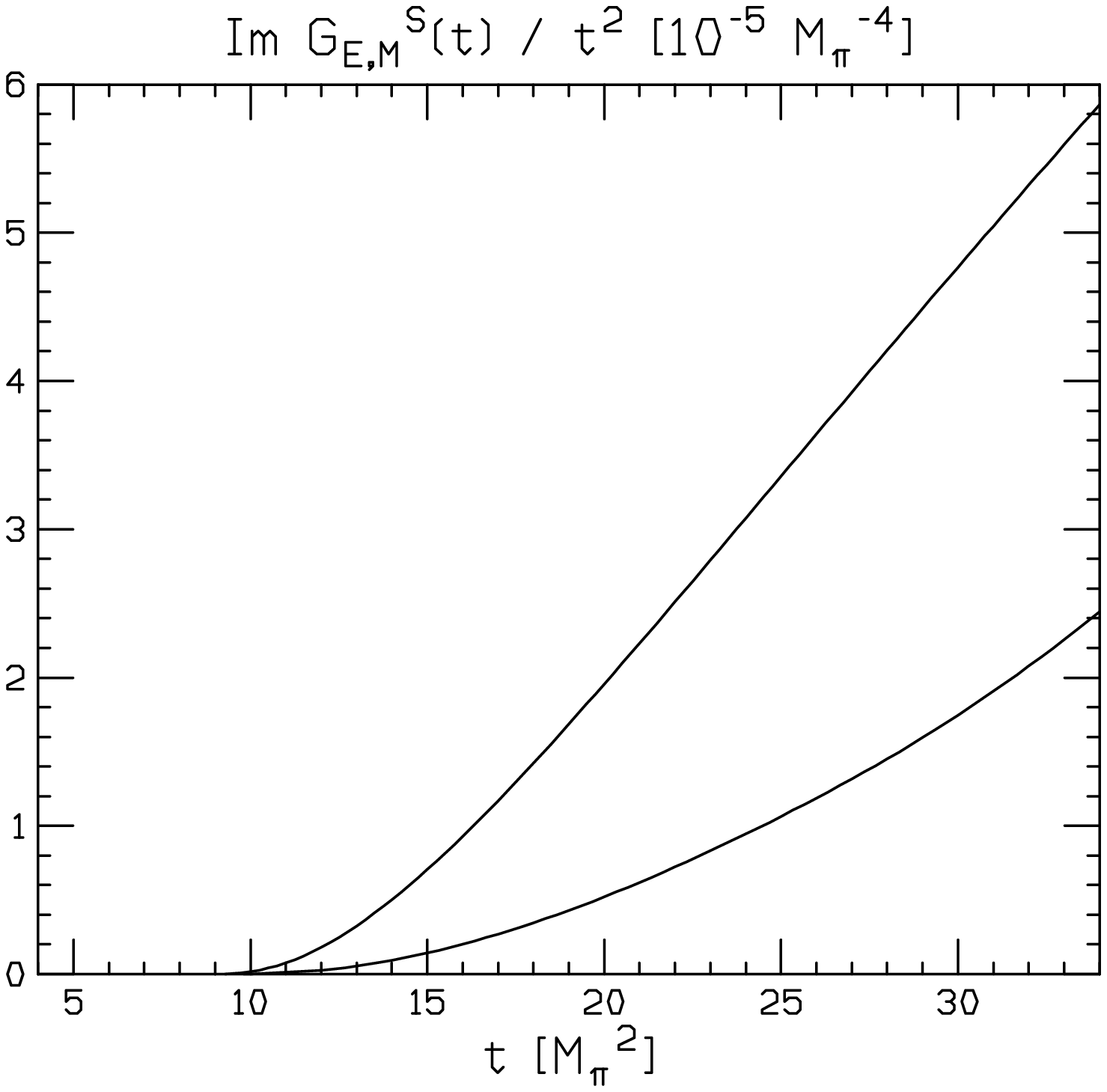}


\vskip -2in
\epsfysize=2.in
\hskip .5truein
\epsffile{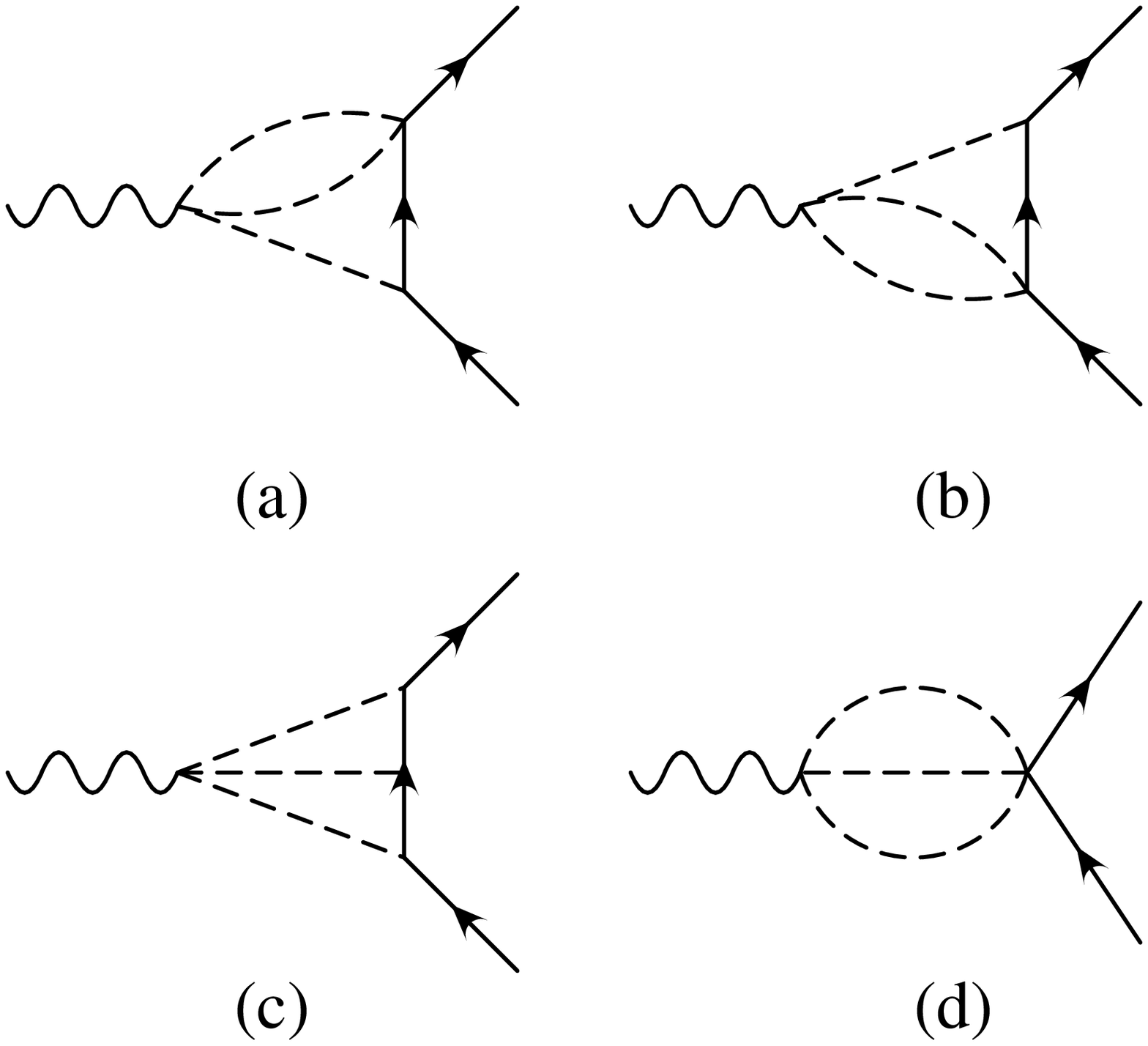}
\vspace{-0.1cm}
 \caption{
      Isoscalar spectral functions weigthed with $1/t^2$ for the electric
      (lower) and magnetic (upper line) Sachs ff (right panel). 
      In the left panel the underlying two--loop graphs are shown (solid,
      dashed, wiggly lines: Nucleons, pions and photons).
    \label{exfig} }
\end{figure}
Recently, the question whether a similar phenomenon appears in
the isoscalar spectral function has been answered \cite{bkmff}. For that, one has
to consider two--loop graphs as shown in fig.1. Although the analysis of
Landau equations reveals a branch point on the second Riemann sheet, $t_c
= 8.9 \,M_\pi^2$, close to the threshold $t_0 = 9M_\pi^2$,
the three--body  phase factors suppress its influence
in the physical region. Consequently, the spectral functions rise smoothly
up to the $\omega$ pole and the common practise of simply retaining vector
meson poles at low $t$ in the isoscalar channel is justified.
Another constraint comes from the {\it neutron radius}. 
Over the last years, the charge radius of the neutron has been 
determined very accurately by  measuring the neutron--atom
scattering length, i.e. $dF_1^n (Q^2) /d Q^2$ at $Q^2 = 0$. This value, 
which we take from the recent paper  \cite{kop}, has to be imposed 
as a boundary
condition on the pertinent spectral functions. Furthermore, 
{\it pQCD} at large $-t$ in its most simple fashion leads to the so--called
dimensional counting, i.e. the large--$Q^2$ fall--off behaviour of the 
ffs is given by the number of constituents and additional spin--flip
suppression factors. For the dispersion relations,  
this leads to a set of superconvergence relations for Im~$F_1 (t)$, 
Im~$F_2 (t)$ and  Im~$t \, F_2 (t)$, which have to be imposed 
($F_2$ is suppressed by one more power in $t$ than $F_1$ due to 
the spin--flip). Logarithmic corrections due to the QCD evolution can also 
be implemented. The simplest (but not unique) way to build  these in the
spectral functions is by means of a logarithm,
\beq
L(t) \equiv \biggl[ \ln \biggl( \frac{\Lambda^2 - t}{Q_0^2} \biggr) 
\biggr]^{-\gamma} \quad ,
\eeq
where $\gamma =2.15$ is the anomalous dimension, $Q_0 \sim \Lambda_{QCD}$ and
the parameter $\Lambda$ can be considered as a boundary between the 
hadronic and the quark phase. It thus signals the onset of pQCD. Note also
that asymptotically, pQCD predicts the same values for the ffs in the 
space-- and time--like regions up to a correction of ${\cal O}(\alpha_s)$
\cite{ster}. To summarize, the isovector spectral functions are
completely fixed from $t=(4 \ldots 50) \,M_\pi^2$ due to unitarity
 (this contribution $F^{2\pi}$ includes the $\rho$ meson).
At large $|t|$, pQCD determines the behaviour of all isovector/isoscalar
spectral functions. In additon, we have a few more isovector and 
isoscalar poles and thus the corresponding fit functions  \cite{mmd}
\beq
F_i^{s,v} (t) = \biggl[ \tilde{F}^{2\pi} (t) \delta_{Iv} +  
\sum_{s,v} \frac{a_i^{s,v} L^{-1} \, (M^2_{s,v})}{M^2_{s,v} -t} 
\biggr] \, L(t) \,, \quad (i=1,2) \, .
\eeq 
I did not yet specify the number of poles. This is were the 
{\it stability criterion} has to be applied.
It states that the number of  meson poles is minimized 
by the requirement that the data can be well fitted, i.e. increasing
this number does not improve the $\chi^2$ any more (for details, 
see Refs.\cite{mmd},\cite{hoeh76}).

\section{Results for the space--like form factors}

\noindent It is instructive to count parameters. Applying the stability
criterion leads to three isocalar and three isovector poles. We thus
have 19 free parameters (6 masses, 12 residua and $\Lambda/Q_0$). The ff
normalization conditions and superconvergence relations
 together with the slope
of $F_1^n$ lead  to 11 constraints, leaving 8 free parameters.
In contrast to a previous dispersive analysis \cite{hoeh76}, 
we are able to identify all three
isoscalar poles  with physical ones,
${S'} = \omega$,  ${S''} = \phi$, ${S'''}= \omega(1600)$ or $\omega(1420)$
(both options are equally viable and lead to the same results, with the
exception of some strangeness components, see below),
and similarly for two of the isovector ones, $\rho' (1450)$ and 
$\rho'' (1690)$. Only the third isovector mass is so tightly fixed 
by the constraints that it can not be chosen
freely. This leaves us with just three fit parameters. The best fit
to the nucleon form factors is shown in Fig.1 (to be precise, we show the
ffs normalized to the dipole fit, $G_D (Q^2) = 
(1+Q^2/.71\,{\rm GeV}^2)^{-2}$. 
In case of $G_E^n$ we normalize to the Saclay data \cite{pla} 
for the Paris potential with the parameters adjusted
to give the exact radius). For a review on the present status of
extracting these form factors, see Klein's talk \cite{fritz}.
\begin{figure}[bht]
\hskip 1.4in
\epsfysize=4.8in
\epsffile{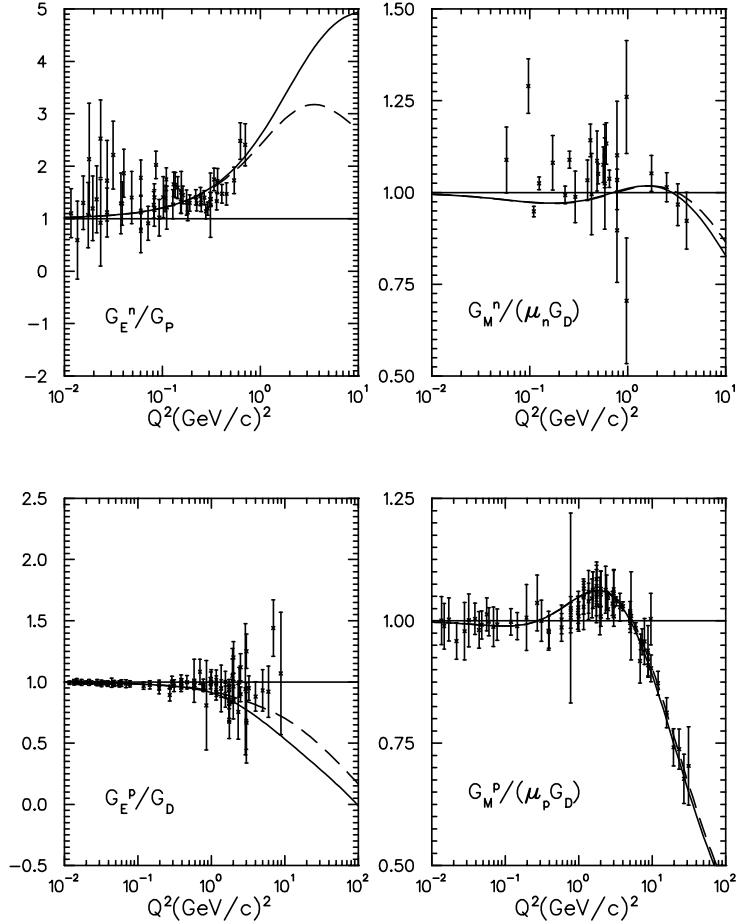}
\vspace{-0.1cm}
 \caption{
    Best fit to the nucleon em form factors. 
    Dashed lines: Space-like data only. Solid lines: Inclusion of the
    time-like data as given before this workshop.} 
\end{figure}
%
%
From these, we deduce the following nucleon electric (E) and 
magnetic (M) radii  \cite{mmd}:
\beq
r_E^p = 0.847~{\rm fm} \, , \, \,
r_M^p = 0.853~{\rm fm} \, , \, \,
r_M^n = 0.889~{\rm fm} \, , \, \,
\eeq
all with an uncertainty of about 1\%. These results are similar to the
ones found by H\"ohler et al.\cite{hoeh76} with the exception of $r_M^n$ which 
has increased by 5\% (due to the neglect of one superconvergence relation in
Ref.\cite{hoeh76}). From the residua at the two lowest isovector poles, 
we can determine the $\omega NN$ and $\phi NN$ coupling constants,
\beq \label{coup}
\frac{g_{\omega NN}^2}{4\pi} = 34.6 \pm 0.8 \, , \,\, \, \kappa_\omega 
= -0.16 \pm 0.01 \, , \, \, 
\frac{g_{\phi NN}^2}{4\pi} = 6.7 \pm 0.3 \, , \,\, \, \kappa_\phi 
= -0.22 \pm 0.01 \, ,
\eeq
where $\kappa_V$ ($V = \omega, \phi$) is the tensor--to--vector coupling 
strength ratio. These results are similar to the ones in
Ref.\cite{hoeh76}. The $\phi$ couplings will be discussed in more detail 
below.  

Of particular interest is the 
onset of pQCD. Only for $G_M^p(t)$ data for
$Q^2 > 10$~GeV$^2$ exist. While these data are consistent with the pQCD
scaling $L^{-1}(Q^2) Q^4 G_M^p(Q^2) \to \, \,  $constant, where $L^{-1}(Q^2)$
accounts for the leading logs, they are not precise enough to rule out a 
non--scaling behaviour, compare Fig.5 in ref.\cite{mmd}. 
Also shown in that figure
is the same quantity without the log corrections.
 All data for the much discussed quantity $Q^2 F_2^p
(Q^2) / F_1^p (Q^2)$ are below $Q^2 = 10$~GeV$^2$ which in our approach is
still in the hadronic region since $\Lambda^2 \simeq 10$~GeV$^2$ for the 
best fit. Stated differently, the tails of the vector meson poles are
still sizeable at $Q^2 \simeq \Lambda^2$ and thus the 
onset of pQCD can not be expected in this regime.

\section{Inclusion of the time--like data}
\noindent In the time--like region, the form factors 
can be determined either in 
$\bar p p$ annihilation  or in
$e^+ e^- \to \bar p p, \bar n n$ collisions \cite{dataold}.
In particular, the FENICE experiment \cite{datfenice} has for the first time
measured the (magnetic) neutron form factor. 
These data and the corresponding ones for the
proton, complemented by the total cross section measurement of $e^+ e^- \to
mh$ below the threshold, seem to indicate a narrow structure at 
$\sqrt{t} = 1.85\,$GeV, as discussed in detail by Voci 
at this workshop \cite{fen}. 
For a comprehensive summary of the status before, see Ref.\cite{baldini}.
It is important to add the following remarks on the extraction of the 
time--like form factors to be discussed. At the nucleon--anti-nucleon 
threshold, one has only S--wave production, consequently
\begin{equation}
G_M (4 m^2) = G_E (4 m^2) \quad .
\label{Gthr}
\end{equation}
Furthermore, at large momentum transfer one expects the magnetic form factor to
dominate. From the data, one can not separate $|G_M|$ from $| G_E|$ so
one has to make an assumption, either setting $|G_M|= | G_E| = |G|$ or
$| G_E|= 0$. Most recent data are presented for the magnetic form factors
\cite{baldini}  and it thus natural to proceed accordingly in the
dispersive analysis, i.e. to  fit the magnetic form factors in the 
time--like region. 
In fig.3, we show the fit to the available space-- and time--like data
available before the workshop\cite{hmd2} for $t \le 6\,$GeV$^2$. At larger
momentum transfer, the unphysical cut due to the logaritm at $t_\Lambda =
\Lambda^2-Q_0^2$ starts to distort the time--like ffs. This should be 
eventually overcome by choosing another function that allows to implement
the leading QCD corrections to the power law fall--off of the ffs. 
These fits are optimized by keeping the  masses of two isovector poles
on physical values and by letting $\Lambda$ vary. This leads to
 $\Lambda^2 = 12.0\,$GeV$^2$, slightly larger than if one fits the space--like
data alone. In that
figure, we also show the result with an additional isoscalar 
pole at 1680 MeV \cite{hmd2}.
Clearly, the trend of the proton data can be described, although the sharp
rise close to threshold is underestimated. In contrast, the magnitude of
the neutron data can not be explained by our three--pole fits. This is
further quantified in fig.4. There, one of the isovector poles is forced
to be at 1850 MeV and the highest isoscalar one at 1600 MeV. The
proton magnetic ff is almost unchanged and the neutron ff still
rises with decreasing $t$, in contrast to the trend of the new FENICE
data. Interestingly, the electric neutron ff (shown by the dashed
line) has a zero close to threshold.
It is important to stress that these data are still too scarce and imprecise
to have an essential effect on the total $\chi^2$ of the fit. Even if one
decreases the empirical uncertainties to tiny numbers, the three--pole fit
can not be forced to have a vanishing neutron form factor at the 
nucleon--anti-nucleon threshold. Finally, I mention that the values for
the nucleon radii and meson couplings given above are not affected by the
inclusion of the time--like data. For a more phenomenological
approach, see ref.\cite{dzsz}.
\begin{figure}[ht]
\begin{minipage}[ht]{77mm}
\epsfxsize=7cm
\epsffile{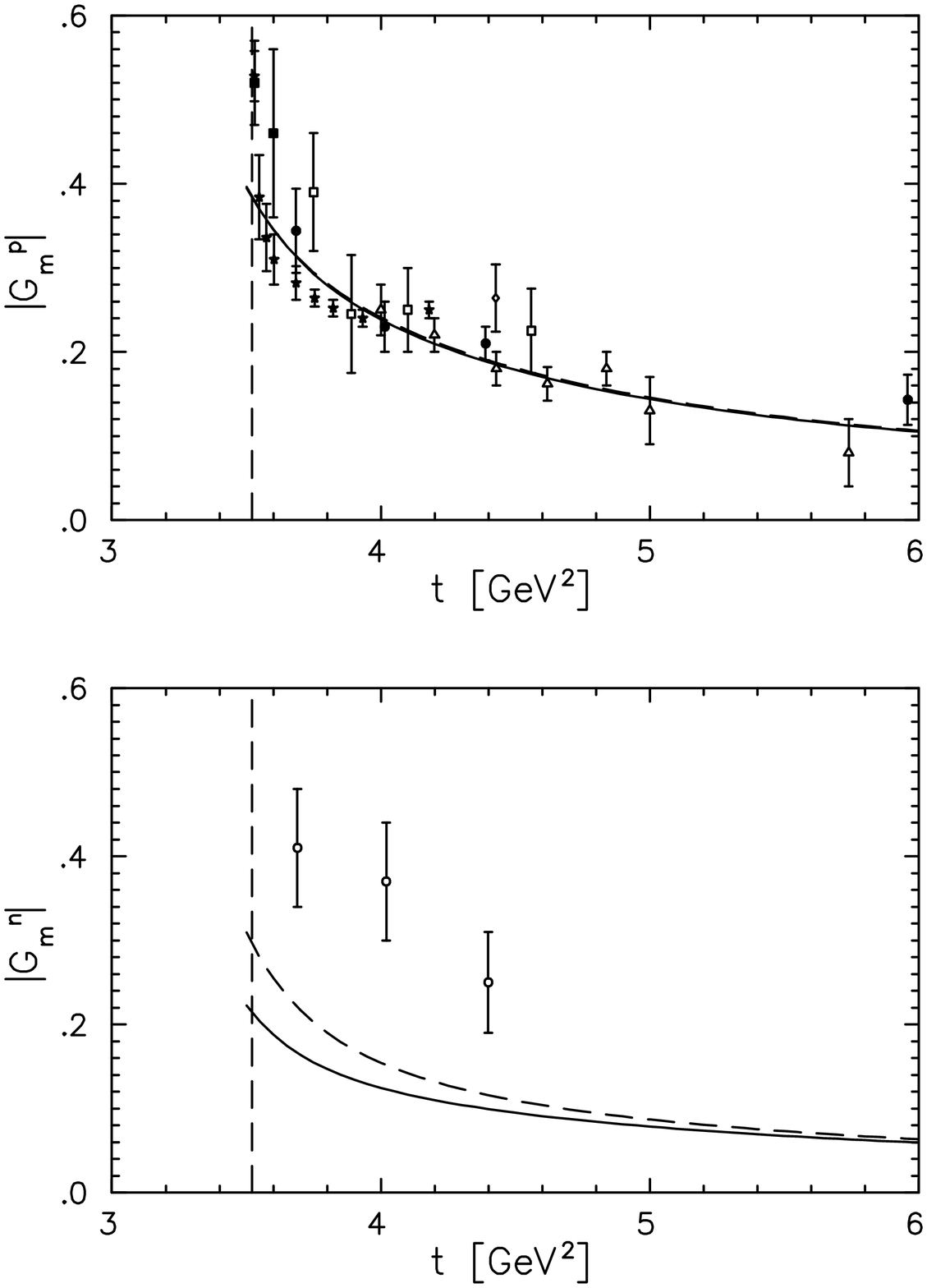}
\caption{Fit to the space-- and time-like data with three (solid lines)
and four (dashed lines) isoscalar poles as described in the text.}
\label{fig:largenenough}
\end{minipage}
\hspace{\fill}
\begin{minipage}[ht]{77mm}
\epsfxsize=7cm
\epsffile{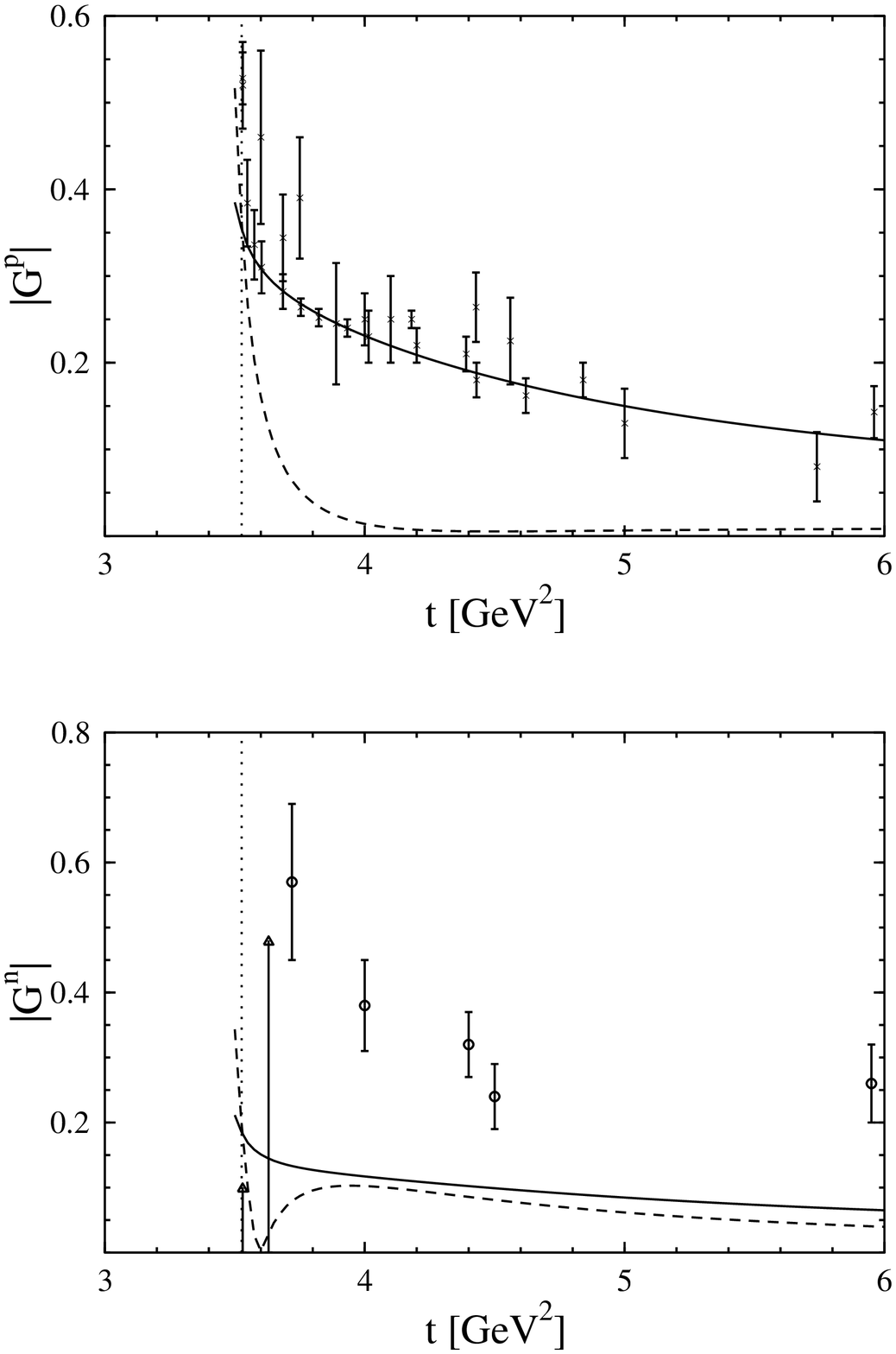}
\caption{Fit to the space-- and time-like data with three is/iv poles.
        One isovector pole fixed at 1.85 GeV as described in the text.
        Dashed lines: Electric ff.}
\label{fig:toosmall}
\end{minipage}
\end{figure}

\section{Strangeness in the nucleon}
\label{sec:strange}

\noindent The isoscalar electromagnetic ffs contain some information 
about the matrix elements of the strange vector current $\bar s \gamma_\mu
s$ in the nucleon. This is most easily seen in the vector dominance model,
where one has a photon--hadron coupling mediated by the $\phi$ meson,
which is almost entirely an $\bar s s$ state. On the other hand, the OZI
rule lets one expect that such a contribution is small since the nucleon is
made of non--strange constituent quarks. In the following, I will discuss two
very different approaches. The first one amounts to a strong violation of
the OZI rule whereas in the second, this rule is respected. The corresponding
strange matrix elements are very different in size, showing that we have not
yet reached a very deep theoretical understanding of this issue.

\subsection{Maximal OZI violation}

\noindent The large $\phi$ meson coupling to the nucleon given in 
eq.(\ref{coup}) was already observed in the fits of the Karlsruhe--Heksinki
group and the consequences concerning the validity of the OZI rule were
discussed in \cite{geho}. 
Jaffe \cite{bob} has shown how one can get bounds on the strange vector
form factors in the nucleon from such dispersion theoretical results. 
It amounts to separating the strangeness contribution from the three isoscalar
poles. For the $\omega$ and the $\phi$, this is done naturally by
observing that there is a small mixing angle with respect to the nonet
flavor eigenstates. The strangeness component of the third pole is
fixed by demanding normalization conditions like $F_1^s (0)=0$ and
conditions on the large momentum fall-off of $F_{1,2}^s (t)$ as 
$t \to -\infty$. The
main assumption of this approach is that these strange form factors have the
same large--$t$ behaviour as the non--strange isoscalar ones. If the fall--off
for the strange form factors is faster, the strange matrix elements will be
reduced. Using the best fit together with a better treatment of the symmetry
breaking in the vector nonet, it is
straightforward to  update Jaffe's analysis. One finds for the
strange magnetic moment and the strangeness radius \cite{hmd} (see
also ref.\cite{fork}),
\beq
\mu_s = -0.24  \, \, {\rm n.m.} \, , \, \, \,
r_s^2 = 0.17 \ldots 0.21   \, \, {\rm fm}^2 \, ,
\eeq
for the fit with the third isoscalar pole at 1.42~GeV (smaller value)
and at 1.6~GeV (larger value for the strange radius). The value for
$\mu_s$ is stable.
Furthermore, the strange ff $F_2^s (t)$ follows a dipole with a cut--off mass
of 1.46~GeV, $F_2^s (t) = \mu_s /(1- t /2.41 \, {\rm GeV}^2)^2$. 
The corresponding strange ffs are shown in fig.\ref{fig:ffslarge}. It is 
important to stress that these numbers should be considered as upper bounds.
Such a strongly coupled $\phi$ meson subsumes other effects like
the strong $\pi\rho$ correlations and it is therefore mandatory to consider
an alternative scenario, in which strength is moved from the $\phi$.

\subsection{OZI resurrected}

\begin{figure}[ht]
\begin{minipage}[ht]{77mm}
\epsfxsize=6cm
\epsffile{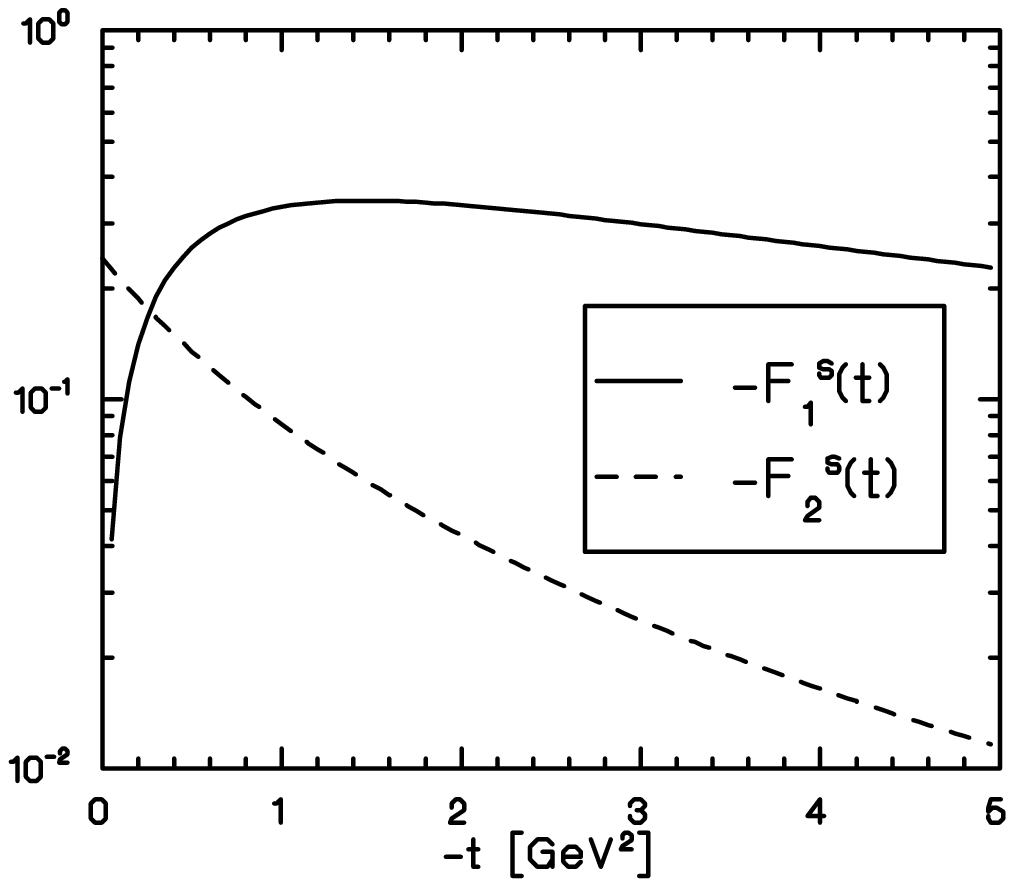}
\caption{Strange ffs based on the approach with maximal OZI violation.}
\label{fig:ffslarge}
\end{minipage}
\hspace{\fill}
\begin{minipage}[ht]{77mm}
\epsfxsize=7cm
\epsffile{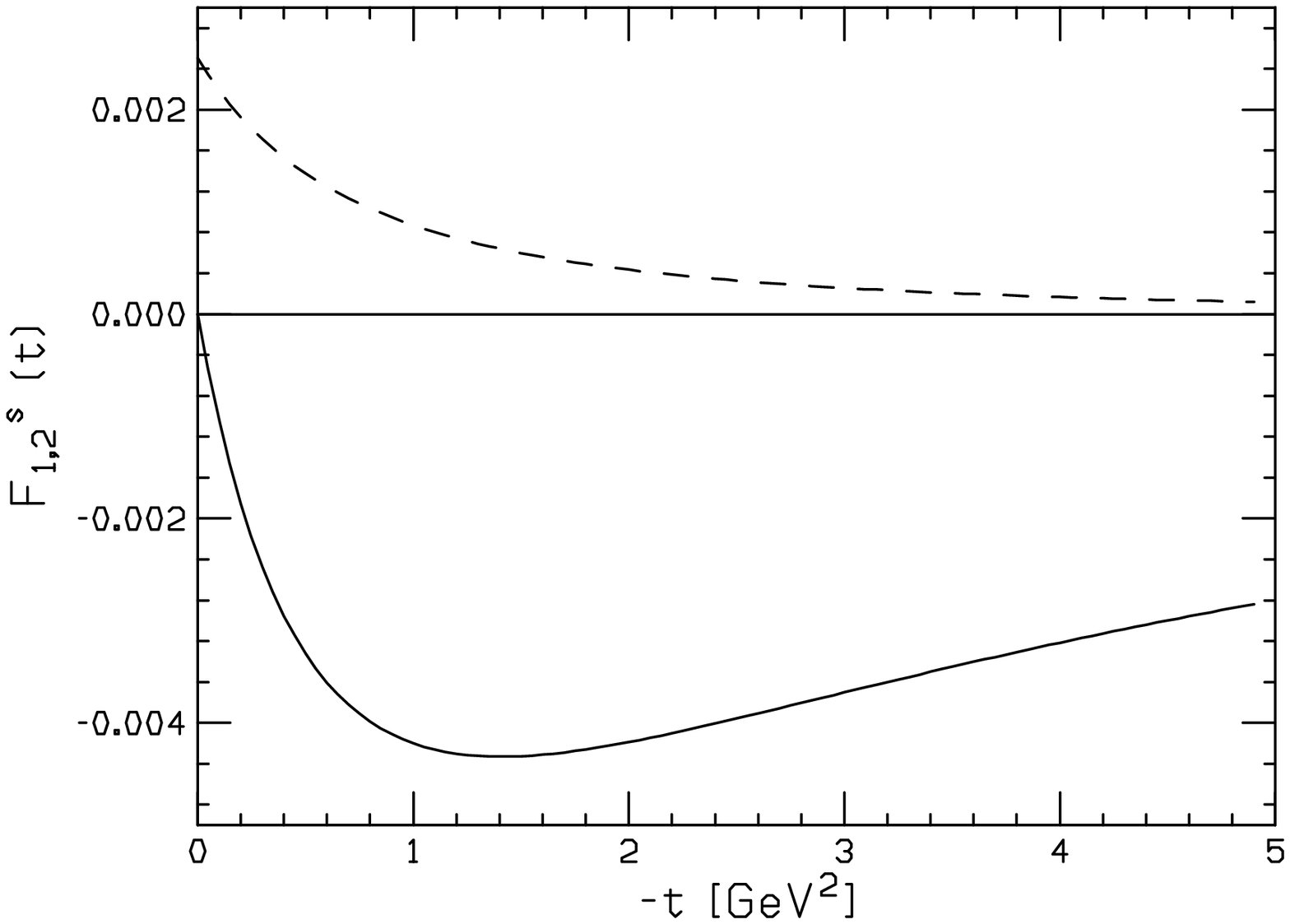}
\caption{Strange ffs based on the approach with the OZI rule 
         imposed. Notice the different y-scale compared to fig.5.}
\label{fig:ffssmall}
\end{minipage}
\end{figure}

\noindent The $\phi$ meson has a sizeable branching fraction into 
the $\pi \rho$
final state. Furthermore, it is well known that in meson exchange
models there is an important contribution from the $\pi \rho$
continuum to the 3--pion ($\omega$) exchange. In the Bonn--J\"ulich
Potential, this contribution has been calculated in ref.\cite{jans}. There,
a parametrization of the corresponding $\pi \rho$ spectral function with a mass
of $M_{\omega '} = 1.22\,$GeV and fixed coupling constants was given.
This approach has further been extended to include kaon loops and
hyperon excitations, with the parameters fixed from a study of the
reactions $p \bar p \to \Lambda \bar \Lambda$ and $p \bar p \to 
\Sigma \bar \Sigma$. There are sizeable cancellations between the various
contributions from graphs with
intermediate $K$'s, $K^*$'s and diagrams with the direct hyperon
interactions \cite{mull} leading to a very small $\phi$ coupling,
\beq \label{coup2}
\frac{g_{\phi NN}^2}{4\pi} \simeq 0.005 \, , \,\, \, \kappa_\phi
\simeq \pm 0.2 \,\, .
\eeq
The sign of the tensor coupling is very sensitive to the details of
the calculation. Neglecting the $\omega\,\phi$ mixing, one can now
perform a fit with the OZI rule imposed. For that, one takes
the poles corresponding to the $\pi \rho$ continuum and the $\omega$ and $\phi$
mesons with fixed couplings given by the model and has as remaining free
parameters the mass of the fourth isoscalar pole ($M_S$), the strength
$a_1^S$ and the mass
of the third isovector pole. The other parameters are constrained as
described before (normalizations, superconvergence relations etc). It
is important to stress that one has to include a fourth isoscalar pole
so as to be able to fulfill all these constraints. The corresponding 
strange form factors take the form \cite{msvo}
\beq
F_1^s (t) = t\,L(t)\,a_1^\phi L_\phi^{-1} \, \frac{M_\phi^2 - M_S^2}{(t
- M_\phi^2)(t-M_S^2)}\, , \,\,
F_2^s (t) = L(t)\,a_2^\phi L_\phi^{-1} \, \frac{M_\phi^2 - M_S^2}{(t
- M_\phi^2)(t-M_S^2)}\, , \,\,
\eeq
imposing the same large-$t$ constraints as for the isoscalar
ffs and $ L_\phi^{-1} = 1/L(M_\phi^2)$. Clearly, the size of these
strange ffs is given by the strength
of the $\phi$--nucleon couplings (as encoded in the residua $a_{1,2}^\phi$).
For a typical set of coupling constants, we find (for a more detailed
account, see ref.\cite{msvo}) 
\beq
\mu_s = 0.003 \, \, {\rm n.m.} \, , \, \, \,
r_s^2 = 0.002   \, \, {\rm fm}^2 \, ,
\eeq
which are two orders of magnitude smaller than the numbers based on
the ansatz with maximal OZI violation. The corresponding strange 
ffs are shown in fig.\ref{fig:ffssmall}. Including the $\omega\,\phi$
mixing would lead to somewhat larger values but not change the
conclusion that the extraction of the strange vector current matrix
elements from a dispersion-theoretical fit to the em ffs hinges
strongly on whether one views the $\phi$ as an isolated pole with a
large effective coupling or whether one shifts a large amount of the
strength from the region around $t \simeq 1\,$GeV$^2$ into a pole that
parametrizes the strong $\pi \rho$--correlations seen in the analysis
of the NN interaction. 

\subsection{Strangeness and unitarity}

\noindent A different approach to  get bounds on the matrix elements
of the strange vector current $\bar s \gamma_\mu s$ has been discussed
in ref.\cite{muhd}. Dispersion relations are used to study the
contribution from the $\bar K K$ intermediate state to the ff spectral
functions. As expected, a direct calculation shows that using the $K
\bar K \to N \bar N$ amplitudes in the Born (tree) approximation in
the dispersion relations is equivalent to a one-loop calculation
within the effective field theory approach. For the latter, the SU(3)
non--linear $\sigma$ model is chosen. A serious violation of unitarity
is found if one translates the bounds on the helicity amplitudes 
for the $KN$ scattering amplitude into bounds for the spectral
functions. This means that resonant and non--resonant kaon
rescattering (or higher loop effects) can not be neglected in such
type of analysis. This agrees with the findings of the model discussed
in the previous paragraph. Furthermore, it is pointed out in
\cite{muhd} that the effect of the strange kaon ff defined via
\beq
\langle 0| \bar s \gamma_\mu s| K^- (k_1) K^+ (k_2)\rangle = (k_1 -
k_2)_\mu \, F^s_K (t) \,\,\, ,
\eeq
is non--negligible if e.g. $F^s_K (t)$ is modelled by a
Gounaris--Sakurai form peaked around the mass of the $\phi$. A refined
analysis which uses the empirical information on $KN \to KN$ or $K\bar
K \to N \bar N$ data continued appropriately to
get more stringent bounds on the strange
radius and magnetic moment is possible but not yet available.

\section{Outlook}

\noindent To my opinion, there are three major issues to be resolved.

\begin{itemize}

\item[$\star$] Better and more consistent data in the space--like
region for low, intermediate and large momentum transfer 
are necessary to further sharpen the values of the
nucleon radii and vector meson coupling constants. This is an
experimental problem. The feasibility of improving upon the existing
data basis is discussed in Klein's talk \cite{fritz}.

\item[$\star$] More theoretical work is needed to get a better handle
on the matrix elements of the strange vector current, as exemplified
in Sec.~\ref{sec:strange} by the approach which accounts for the
strong $\pi \rho$ correlations and the bounds from unitarity. A deeper
theoretical understanding is necessary for setting the stage for the
strange form factor measurements at TJNAF.

\item[$\star$] The neutron ffs in the time--like region have to be
measured more precisely close to the nucleon--anti-nucleon threshold.
If the interesting structure indicated by the FENICE data persists,
theory is challenged to explain it. This might finally be the trace of
the much searched for dibaryon states.

\end{itemize}

Finally, it is important to stress that all these problems are
intertwined. For example, the extraction of the strange ffs from
parity--violation experiments can only be done precisely if the data
are accurate enough but also the non--strange ffs are known precisely,
since in most parity--violation experiments the latter are used as
amplification factors.

\section*{Acknowledgements}

It is a pleasure to thank the organizers, in particular Prof. Baldini,
for their invitation and kind hospitality. I am grateful to
Prof. Baldini for supplying me with the new and updated FENICE results
prior to publiation, to Hans--Werner Hammer for providing me with some
novel results and to Josef Speth and Wally van Orden for allowing me
to present material before publication. Last but not least Dieter
Drechsel is thanked for constant support and Nathan Isgur for some
pertinent comments.


\begin{thebibliography}{99}

\bibitem{lamb}M. Weitz  et al., Phys. Rev. Lett. 72 (1994) 328;
D.J. Berkeland et al., Phys. Rev. Lett. 75 (1995) 2470

\bibitem{frafu} W.R. Frazer and F.J. Fulco,  Phys. Rev. Lett. 
2 (1959) 365; Phys. Rev. 117 (1960) 1609

\bibitem{na} Y. Nambu, Phys. Rev. 106 (1957) 1366

\bibitem{sa} J.J. Sakurai, Ann. Phys. (NY) 11 (1960) 1

\bibitem{pQCD} S.J. Brodsky and G. Farrar,
Phys. Rev. D11  (1975) 1309

\bibitem{fen} C. Voci, these proceedings; R. Baldini, private communication 

\bibitem{gss} J. Gasser, M.E. Sainio and A. ${\rm \check S}$varc, 
Nucl. Phys. B307 (1988) 779

\bibitem{bkmff} V. Bernard, N. Kaiser and Ulf-G. Mei\ss ner,
[hep-ph/9607428], Nucl. Phys. A, in print

\bibitem{kop} S. Kopecky et al.,  Phys. Rev. Lett. 
74  (1995) 2427

\bibitem{ster} L. Magnea and G. Sterman, Phys. Rev. D42 (1990) 4222

\bibitem{mmd} P. Mergell, Ulf-G. Mei\ss ner and D. Drechsel,
 Nucl. Phys. A596 (1996) 367

\bibitem{hoeh76} G. H\"ohler et al., Nucl. Phys. B114 (1976) 505

\bibitem{pla} S. Platchkov et al., Nucl. Phys. A510 (1990) 740

\bibitem{fritz} F. Klein, these proceedings

\bibitem{dataold}
%
M. Castellano et al., Nuovo Cim. A14 (1973) 1;
G. Bassompierre et al., Phys. Lett. B64 (1976) 475, B68 (1977) 477,
Nuovo Cim. A73 (1983) 347;
B. Delcourt et al., Phys. Lett. B86 (1979) 395;
D. Bisello et al., Nucl. Phys. B224 (1983) 379; 
G. Bardin et al., Phys. Lett. B255 (1991) 149; B257 (1991) 514;
Nucl. Phys. B411 (1994) 3

\bibitem{datfenice} A. Antonelli et al., Phys. Lett. B313 (1993) 283;
B334 (1994) 431

\bibitem{baldini} R. Baldini and E. Pasqualucci, in "Chiral Dynamics:
Theory and Experiment", A.M. Bernstein and B.R. Holstein (eds.),
Springer, Heidelberg, 1995

\bibitem{hmd2} H.--W. Hammer, Ulf-G. Mei\ss ner and D. Drechsel,
 Phys. Lett. B385 (1996) 343

\bibitem{dzsz} Z. Dziembowski and A. Szczurek, Phys. Lett. B387 (1996)
875

\bibitem{geho} H. Genz and G. H\"ohler, Phys. Lett. B61 (1976) 389

\bibitem{bob} R.L. Jaffe, Phys. Lett. B229 (1989) 275

\bibitem{hmd} H.--W. Hammer, Ulf-G. Mei\ss ner and D. Drechsel,
 Phys. Lett. B367 (1996) 323

\bibitem{fork} H. Forkel, preprint ECT*/Sept/014-95, 1995

\bibitem{jans} G. Jan{\ss}en, Dissertation, University of Bonn, 1995

\bibitem{mull} V. Mull, Dissertation, University of Bonn, 1996

\bibitem{msvo} Ulf-G. Mei\ss ner, J. Speth and W. van Orden, in preparation

\bibitem{muhd} M. Musolf, H.--W. Hammer and D. Drechsel, preprint [hep-ph/9610402]

\end{thebibliography}
\end{document}